\documentclass[conference]{IEEEtran}
\usepackage{amsmath}
\usepackage{amssymb}
\usepackage{graphicx}
\usepackage{color}
\usepackage{cite}
\usepackage{bm}
\usepackage{epsfig,psfrag}
\usepackage{tabularx}
\usepackage{acronym}
\usepackage[yyyymmdd,hhmmss]{datetime}
\usepackage{bbm}
\usepackage{mathrsfs} 
\usepackage{bardiag}
\usepackage{tabularx}
\usepackage{multirow}
\usepackage{colortbl,pgfplotstable}
\usepackage{LatexInclusion/notation}
\usepackage[level=3]{LatexInclusion/messages}  % set level to 0, to hide all notes
\usepackage{subfig}
\usepackage{enumitem}

\usepackage{auto-pst-pdf}

\newcommand{\bd}{\begin{description}}
\newcommand{\ed}{\end{description}}
\newcommand{\be}{\begin{enumerate}}
\newcommand{\ee}{\end{enumerate}}
\newcommand{\bi}{\begin{itemize}}
\newcommand{\ei}{\end{itemize}}
\newcommand{\bl}{\begin{list}}
\newcommand{\el}{\end{list}}
\newcommand{\bt}{\begin{tabbing}}
\newcommand{\et}{\end{tabbing}}

\definecolor{BLUE}{rgb}{0,0,1}

\allowdisplaybreaks
\definecolor{BLUE}{rgb}{0,0,1}

\providecommand{\bl}{\textcolor{blue}}

\providecommand{\ist}{\hspace*{.3mm}}

\providecommand{\rmv}{\hspace*{-.3mm}}

\providecommand{\nn}{\nonumber}

\acrodef{siso}[SISO]{single-input single-output }
\acrodef{bp}[BP]{belief propagation}
\acrodef{slam}[SLAM]{simultaneous localization and mapping}
\acrodef{2d}[2-D]{two dimensional}
\acrodef{pa}[PA]{physical anchor}
\acrodef{va}[VA]{virtual anchor}
\acrodef{los}[LOS]{line of sight}
\acrodef{pf}[PF]{potential features}
\acrodef{nlos}[NLOS]{non-line of sight}
\acrodef{mmse}[MMSE]{minimum mean square error}
\acrodef{rmse}[RMSE]{root mean square error}
\acrodef{mpc}[MPC]{multipath component}
\acrodef{pdf}[PDF]{probability density function}
\acrodef{cdf}[CDF]{cumulative distribution function}
\acrodef{gospa}[GOSPA]{generalized optimal sub-pattern assignment}
\acrodef{iid}[i.i.d.]{independent and identically distributed}

\newcommand{\T}{\mathrm{T}}
\newcommand{\CH}{\mathrm{H}}
 
\title{
{A Belief Propagation Approach\\
for Direct Multipath-Based SLAM}}
\author{
\IEEEauthorblockN{Mingchao Liang\IEEEauthorrefmark{1}, Erik Leitinger\IEEEauthorrefmark{2}, and Florian Meyer\IEEEauthorrefmark{1} \\[1.2mm]
\IEEEauthorblockA{\IEEEauthorrefmark{1}Department of Electrical and Computer Engineering, University of California San Diego, USA \\
\texttt{\{m3liang, flmeyer\}@ucsd.edu} \\[1mm]
}}
\IEEEauthorblockA{\IEEEauthorrefmark{2}Signal Processing and Speech Communication Laboratory, Graz University of Technology, Austria} 
\texttt{erik.leitinger@tugraz.at}
\vspace*{-3mm}
}

\definecolor{BLUE}{rgb}{0,0,1}

\definecolor{myred}{rgb}{1,0.27,0}
\definecolor{mygreen}{rgb}{0.1, 0.55, 0.1}
\definecolor{myblue}{rgb}{0, 0, 1}

\begin{document}
\maketitle

\begin{abstract}
In this work, we develop a multipath-based \ac{slam} method that can directly be applied to received radio signals. In existing multipath-based \ac{slam} approaches, a channel estimator is used as a preprocessing stage that reduces data flow and computational complexity by extracting features related to \acp{mpc}. We aim to avoid any preprocessing stage that may lead to a loss of relevant information. The presented method relies on a new statistical model for the data generation process of the received radio signal that can be represented by a factor graph. This factor graph is the starting point for the development of an efficient \ac{bp} method for multipath-based \ac{slam} that directly uses received radio signals as measurements. Simulation results in a realistic scenario with a single-input single-output (SISO) channel demonstrate that the proposed direct method for radio-based \ac{slam} outperforms state-of-the-art methods that rely on a channel estimator.
\vspace{1mm}
\end{abstract}

\begin{IEEEkeywords}
Simultaneous localization and mapping, SLAM, multipath channel, belief propagation, and factor graph. \vspace{1mm}
\end{IEEEkeywords}

%-------------------------------------------------------------------------------------

%-------------------------------------------------------------------------------------

\acresetall
\section{Introduction}

Situational awareness in the indoor environment is critical in various applications, including autonomous navigation, public safety, and asset tracking. Multipath-based \ac{slam} is a promising approach to estimating the positions of mobile users and features in the propagation environment. By associating \acp{mpc} in radio signals with the geometry of the radio reflectors, multipath propagation is exploited to increase positioning accuracy and build a partial map of the indoor environment. Conventional multipath-based \ac{slam} methods \cite{LeiMeyHlaWitTufWin:J19, MenMeyBauWin:JS19, LeiGreWit:19,MeyGem:J21,GenJosWan:J16, KadBehSorWelAmiYerYoo:22} use a channel estimator \cite{ShuWanJos:13,GerMecChrXenNan:J16,HanFleuRao:J18,GreLeiWitFle:J23} to detect and extract \acp{mpc} by preprocess received radio signals in blocks of samples or ``snapshots''. The parameters of each extracted \ac{mpc} include delay, angle of arrival, and angle of departure. These parameters are used as noisy measurements for a \ac{slam} method that is based on \ac{bp} \cite{LeiMeyHlaWitTufWin:J19, MenMeyBauWin:JS19,LeiGreWit:19,MeyGem:J21}, Rao-Blackwellized particle filtering \cite{GenJosWan:J16}, or neural networks \cite{KadBehSorWelAmiYerYoo:22}. This two-step processing is widely used as it reduces data flow and, thus, the computational complexity of SLAM. However, important information may be lost in this preprocessing stage. In particular, if parameters of multiple \acp{mpc}  are very similar the channel estimator may detect them as a single \ac{mpc} due to finite resolution capabilities limited by signal bandwidth. This can lead to a significantly degraded \ac{slam} performance.
\begin{figure}[!tbp]
\vspace{2mm}
	\centering
	% \psfrag{A1}[r][r][1.00]{\raisebox{-0mm}{$\hspace{0mm}\V{p}_k$}}
	% \psfrag{B1}[c][c][1.00]{\raisebox{1mm}{$\V{p}_{1}^{(1)}$}}
	% \psfrag{C1}[l][l][1.00]{\raisebox{1mm}{$\hspace{1mm} \V{p}_{2}^{(1)}$}}
	\psfrag{CE1}[c][c][1.00]{Channel}
	\psfrag{CE2}[c][c][1.00]{\hspace{-.5mm} Estimator}
	\psfrag{MPC1}[c][c][1.00]{Estimated}  
	\psfrag{MPC2}[c][c][1.00]{MPCs}
	\psfrag{SL1}[c][c][1.00]{SLAM}
	\psfrag{SL2}[c][c][1.00]{\hspace{1mm}Direct SLAM}
	\psfrag{R1}[c][c][1.00]{Radio Signal}
	\psfrag{T1}[l][l][1.00]{Previous Work}
	\psfrag{T2}[l][l][1.00]{\color{red}{\bf Proposed}}
	\psfrag{RE11}[l][l][1.00]{Estimated Agent}
	\psfrag{RE12}[l][l][1.00]{Position $\hat{\V{p}}_k$}
	\psfrag{RE13}[l][l][1.00]{Estimated Feature}
	\psfrag{RE14}[l][l][1.00]{Positions $\hat{\V{p}}_{k, n}^{(j)}$}
	\psfrag{RE21}[l][l][1.00]{Estimated agent}
	\psfrag{RE22}[l][l][1.00]{position $\hat{\V{p}}_k$}
	\includegraphics[scale=0.8]{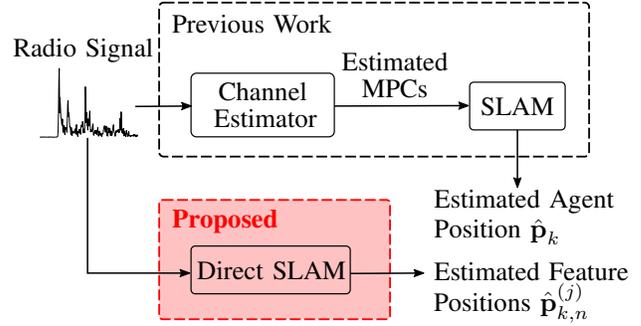}
	\vspace{2mm}
	\caption{Flow diagram of the proposed direct multipath-based SLAM that use the received radio signal as measurements. A flow diagram of conventional multipath-based SLAM that uses the estimated \acp{mpc} provided by the channel estimator as measurements is also shown. }
	\label{fig:diagram}
	\vspace{-4mm}
\end{figure}

In this paper, we propose a multipath-based \ac{slam} method for \ac{siso} systems that avoids any preprocessing stage by directly using received radio signals as measurements. Our approach addresses direct multipath-based \ac{slam} as a joint sequential Bayesian inference problem. Fig.~\ref{fig:diagram} shows the flow diagram of the proposed direct method compared to existing methods with a channel estimator. 

For direct SLAM, we introduce a new statistical model to describe the data-generating process of received radio signals, which is a combination of the Swerling 1 model for correlated measurements in \cite{LepRabGla:J16} and Bernoulli existence model in \cite{LiaKroMey:J23}. This statistical model can be represented well by a factor graph \cite{Loe:04,KscFreLoe:01}. Based on the factor graph, an efficient sequential \ac{bp} message-passing method for the estimation of mobile agent position and features in the propagation environment is developed. For an accurate approximation of BP messages, following our previous work in \cite{MeyHliHla:J16,MeyBraWilHla:J17,LiaKroMey:J23}, we represent some of the \ac{bp} messages by random samples ``particles'' and others by a mean and covariance matrix obtained via moment matching. The main contributions of this work are summarized as\vspace{0mm} follows.
\begin{itemize}
    \item We introduce a new statistical model for multipath-based SLAM using received radio signals.
    \vspace{1.5mm}
    
    \item We develop an efficient \ac{bp} method for direct multipath-based \ac{slam}.
    \vspace{1.5mm}
\end{itemize}
By comparing the proposed direct approach with a state-of-the-art reference method for multipath-based \ac{slam}, we demonstrate that directly using the received radio signal as a measurement can lead to an improved \ac{slam}\vspace{0mm} performance. 
% \vspace{1.5mm}

% The proposed method is based on the \ac{bp} algorithm \cite{Loe:04} and a statistical model for the \rd{???} .

%-------------------------------------------------------------------------------------

%-------------------------------------------------------------------------------------
\section{System Model} \label{sec:sys} 

\begin{figure}[!tbp]
    \centering
    \psfrag{A1}[r][r][0.80]{\raisebox{-0mm}{$\hspace{0mm}\V{p}_k$}}
    \psfrag{B1}[c][c][0.80]{\raisebox{1mm}{$\V{p}_{1}^{(1)}$}}
    \psfrag{C1}[l][l][0.80]{\raisebox{1mm}{$\hspace{1mm} \V{p}_{2}^{(1)}$}}
    \psfrag{W1}[l][l][0.80]{Reflecting Surface}
    \psfrag{W2}[l][l][0.80]{LOS Path}
    \psfrag{W3}[l][l][0.80]{NLOS Path}
    \psfrag{W4}[l][l][0.80]{Agent}
    \psfrag{W5}[l][l][0.80]{VA}
    \psfrag{W6}[l][l][0.80]{PA}
    \includegraphics[scale=0.9]{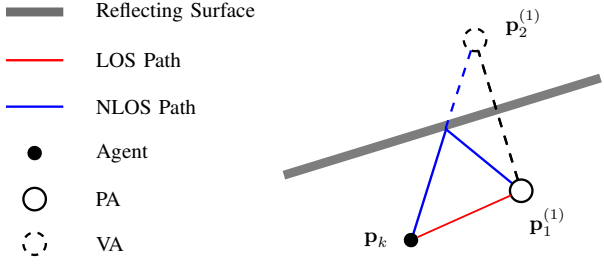}
    \vspace{1.5mm}
    \caption{ Scenario with one mobile agent, one physical anchor (PA), and one virtual anchor (VA).}
    \label{fig:va_example}
    \vspace{-4mm}
\end{figure}

For multipath-based \ac{slam}, we consider a mobile agent with an unknown time-varying position $\V{p}_k \in \mathbb{R}^2$ and $J$ \acp{pa} with known positions $\V{p}_{1}^{(j)} \in \mathbb{R}^2, j \in \{1, \dots, J\}$, where $k$ is the index of discrete time. The number of physical anchors, $J$, is assumed to be known. There are $L_k^{(j)} \rmv-\rmv 1$ \acp{va} with unknown position $\V{p}_{l}^{(j)} \in \mathbb{R}^2, l = \{2, \dots, L_k^{(j)}\}$ associated to the $j$-th \ac{pa}. The number of VAs, $L_k^{(j)}$, is time-varying and unknown. VAs are the mirror images of the \ac{pa} at reflecting surfaces. Fig.~\ref{fig:va_example} shows a scenario with one \ac{pa} and one \ac{va}.  
%\acp{pa} and \acp{va} are geometric features that characterize the propagation environment. 

The number of \acp{va} $L_k^{(j)}$ is time-varying and unknown. To address this, at each discrete time step $k$, we introduce \ac{pf} \cite{MeyKroWilLauHlaBraWin:J18} indexed by $(j, n), j \in \{1, \dots, J\}, n \in \{1, \dots, N_k^{(j)}\}$. The existence of each \acp{pf} is modeled by binary random variables $r_{k, n}^{(j)} \in \{0,1\}, j \in \{1, \dots, J\}, n \in \{1, \dots, N_k^{(j)}\}$. Here $N_k^{(j)}$ is the maximum possible number of features at time step $k$, i.e., $N_k^{(j)} \geq L_k^{(j)}$. The state of a \ac{pf} is denoted by $\V{y}_{k, n}^{(j)}$, which includes their position $\V{p}_{k, n}^{(j)}$, existence $r_{k, n}^{(j)}$, and intensity $\gamma_{k, n}^{(j)}$. We further introduce the notation $\V{\phi}_{k, n}^{(j)} = [\V{p}_{k, n}^{(j) \T} \gamma_{k, n}^{(j)}]^\T$. Note that \acp{pa} are also represented by \acp{pf} since their intensity is also time-varying, and \ac{los} paths to PAs may be unavailable. The state of the mobile agent is denoted as $\V{x}_{k}$. It includes the agent's position $\V{p}_k$ and possibly further motion-related parameters. For future reference, we establish the notation $\V{y}^{(j)}_{k} = \big[\V{y}^{(j) \T}_{k, 1} \cdots \V{y}^{(j) \T}_{k, N_k^{(j)}}\big]^\T\rmv\rmv\rmv$.

\subsection{Measurement Model} \label{subsec:sys_radio}

We consider a SISO system, where at each time $k$, the mobile agent transmits a radio signal, which is received by the \acp{pa}. However, note that the proposed model can be easily reformulated for the case where the \acp{pa} act as transmitters and the mobile agent act as a receiver. Let $H(f)$ be the frequency-domain representation of the transmitted radio signal in the baseband. The total bandwidth of the signal is denoted as $B$. The radio signal received by the $j$-th \ac{pa} can now be modeled as\vspace{0mm} \cite[Ch. 2]{Mei:14} 
\begin{equation}
    \vspace{-0mm} Z_k^{(j)}(f)  = \sum_{l = 1}^{L_k^{(j)}} \rho_{k, l}^{(j)} \ist\ist H(f) \text{e}^{-j 2\pi f \tau_{k, l}^{(j)} } + \epsilon_{k}^{(j)}(f). \label{eq:radio_signal}
\vspace{-.3mm}
\end{equation}
Here, $\rho_{k, l}^{(j)} \in \mathbb{C}$ and $\tau_{k, l}^{(j)} = \Vert \V{p}_k - \V{p}_{l}^{(j)} \Vert / c$ are the complex amplitude and the delay related to the $l$-th propagation path and $c$ is the speed of light. The complex amplitude $\rho_{k, n}^{(j)}$ is distributed according to a circularly symmetric complex Gaussian \ac{pdf}. Specifically, the absolute value of $\rho_{k, n}^{(j)}$ is Rayleigh distributed and its phase is uniformly distributed over $[0, 2\pi)$. The complex amplitudes $\rho_{k, l}^{(j)}$ and $\rho_{k'\rmv, l'}^{(j')}$ are independent if $(k,l,j) \rmv\neq\rmv (k'\rmv,l',j')$. This amplitude model is also known as \textit{Swerling} 1 \cite{Sko:B00, LepRabGla:J16}. 

The first term on the right-hand side of \eqref{eq:radio_signal} describes the contribution of the \ac{pa} from the \ac{los} path and the contributions of \acp{va} from the \ac{nlos} paths due to specular reflections. The complex additive noise $\epsilon_{k}^{(j)}(f)$ aggregates measurement noise, diffuse multipath components, and specular paths that cannot be resolved with the available bandwidth. The noise is assumed to be a zero-mean, uncorrelated, circularly-symmetric complex Gaussian process \cite{Ric:05, SalRicKoi:J09, LeiGreFleWit:20}.

Sampling the received signal $Z_k^{(j)}(f)$ with frequency spacing $\Delta$ leads to the measurement vector $\V{z}_k^{(j)} = [z_{k, 1}^{(j)} \cdots z_{k, M}^{(j)}]^\T = \big[z_k^{(j)} \big( \frac{-(M - 1)}{2} \Delta \big) \hspace{1mm} \cdots \hspace{1mm} z_k^{(j)} \big( \frac{(M - 1)}{2} \Delta \big) \big]^\T \in \mathbb{C}^M$ with length $M = B / \Delta + 1$.We can now express the signal model in \eqref{eq:radio_signal} in terms of \ac{pf} states $\V{y}_{k, n}^{(j)} = [\V{p}_{k, n}^{(j) \T} \gamma_{k, n}^{(j)} \hspace{.6mm} r_{k, n}^{(j)} ]^\T\rmv\rmv$, i.e.\vspace{-2mm},
\begin{equation}
    \V{z}^{(j)}_{k} = \sum^{N_k^{(j)}}_{n = 1} \ist r_{k, n}^{(j)} \ist\ist \rho_{k, n}^{(j)} \ist \V{h}_{k, n}^{(j)} + \V{\epsilon}^{(j)}_{k} \label{eq:meas_model}
    \vspace{.5mm}
\end{equation}
with $\V{\epsilon}^{(j)}_{k} \rmv=\rmv \big[\epsilon_{k}^{(j)} \big( \frac{-(M - 1)}{2} \Delta \big) \cdots \hspace{1mm} \epsilon_{k}^{(j)} \big( \frac{(M - 1)}{2} \Delta \big) \big]^\T \rmv\in\rmv \mathbb{C}^M$. Here, the vector $\V{h}_{k, n}^{(j)} \in \mathbb{C}^M$ represents the sampled transmit signal, i.e.,
\begin{align}
    \V{h}_{k, n}^{(j)} &= \Big[ H \Big( \frac{-(M - 1)}{2} \Delta \Big) \text{e}^{-j 2\pi \big( \frac{-(M - 1)}{2} \Delta \big) \tau_{k, n}^{(j)} }  \nn \\
    &\hspace*{10mm} \cdots H \Big( \frac{(M - 1)}{2} \Delta \Big) \text{e}^{-j 2\pi \big( \frac{(M - 1)}{2} \Delta \big) \tau_{k, n}^{(j)} } \Big]^\T\rmv\rmv\rmv. \label{eq:hkn}
\end{align} 
Recall that $\tau_{k, n}^{(j)} = \Vert \V{p}_k - \V{p}_{k,n}^{(j)} \Vert / c$. Based on assumptions made above, the complex amplitude $\rho_{k, n}^{(j)}$ is distributed according to $\mathcal{CN}\big(\rho_{k, n}^{(j)}; 0,\gamma^{(j)}_{k,n}\big)$, and the noise $\V{\epsilon}^{(j)}_{k}$ is distributed according to $\mathcal{CN}\big(\V{\epsilon}^{(j)}_{k}; \V{0}, \M{C}_{\V{\epsilon}}^{(j)} \big)$. The noise covariance matrix $\M{C}_{\V{\epsilon}}^{(j)}$ is assumed unknown and is modeled as a random variable. It can be easily verified that the conditional \ac{pdf} $f \big( \V{z}^{(j)}_{k} | \V{x}_k, \V{y}^{(j)}_{k}, \M{C}_{\V{\epsilon}}^{(j)} \big)$ is also zero-mean complex Gaussian with covariance matrix $\M{C}^{(j)}_{k} = \M{C}_{\V{\epsilon}}^{(j)} + \sum^{N_k^{(j)}}_{n = 1} r_{k, n}^{(j)} \gamma_{k, n}^{(j)} \V{h}^{(j)}_{k, n} \V{h}^{(j) \CH}_{k, n}\vspace{2mm}$.
% In addition, the complex random vector $\V{\epsilon}^{(j)}_{k}$, consists of independent and identically distributed entries that follow the circularly symmetric Gaussian \ac{pdf} $\mathcal{CN}\big(0,\sigma_{\epsilon}^2\big).$

\subsection{State-Transition Model}
\label{sec:models}

The evolution of the agent state $\V{x}_k$, \ac{pf} states $\V{y}_{k, n}^{(j)}$ and the noise covariance $\M{C}_{\V{\epsilon}, k}^{(j)}$ are assumed to be independent across $k$, $j$, and $n$ and are described by the state transition \acp{pdf} $f(\V{x}_{k} | \V{x}_{k - 1})$, $f(\V{y}_{k, n}^{(j)} | \V{y}_{k, n - 1}^{(j)}) = f(\V{\phi}_{k, n}^{(j)}, r_{k, n}^{(j)} | \V{\phi}_{k, n - 1}^{(j)}, r_{k, n - 1}^{(j)})$, and $f(\M{C}_{\V{\epsilon}, k}^{(j)} | \M{C}_{\V{\epsilon}, k - 1}^{(j)})$, respectively. Specifically, if legacy \ac{pf} $n \in \{1, \dots, N_{k - 1}^{(j)}\}$ does not exist at $k - 1$, i.e., $r_{k, n - 1}^{(j)} \rmv=\rmv 0$, then it does not exist at $k$ either. The state-transition \ac{pdf} thus reads $f(\V{\phi}_{k, n}^{(j)}, 1 | \V{\phi}_{k, n - 1}^{(j)}, 0) = 0$ and $f(\V{\phi}_{k, n}^{(j)}, 0 | \V{\phi}_{k, n - 1}^{(j)}, 0) = f_{\text{D}}(\V{\phi}_{k, n}^{(j)})$ with $f_{\text{D}}(\cdot)$ being an arbitrary ``dummy'' \ac{pdf}. If \ac{pf} $n$ exists at $k - 1$, i.e., $r_{k, n - 1}^{(j)} \rmv\rmv=\rmv\rmv 1$, then the probability that it also exists at $k$, is $p_{\text{s}}$, known as the survival probability. The state-transition \ac{pdf} reads  $f(\V{\phi}_{k, n}^{(j)}, 1 | \V{\phi}_{k, n - 1}^{(j)}, 1) \rmv\rmv=\rmv\rmv p_{\text{s}}\ist f(\V{\phi}_{k, n}^{(j)} | \V{\phi}_{k, n - 1}^{(j)}) $ and $f(\V{\phi}_{k, n}^{(j)}, 0 | \V{\phi}_{k, n - 1}^{(j)}, 1) \rmv=\rmv (1 - p_{\text{s}}) f_{\text{D}}(\V{\phi}_{k, n}^{(j)})$.

The birth of newly appearing features associated with \ac{pa} $j$ at time $k$, is modeled by a Poisson point process with mean $\mu_{\text{B}}^{(j)}$ and spatial \ac{pdf} $f_{\text{B}}^{(j)}(\V{\phi}| \V{x}_{k})$.  Based on the assumption that the number of newly appearing features is significantly smaller than the number of measurements, $M$, we introduce $M$ new \acp{pf}, one for each measurement $z_{k, m}^{(j)}$ \cite{LiaKroMey:J23}. Thus, $N_{k}^{(j)} = N_{k - 1}^{(j)} + M$ \cite{LiaKroMey:J23}. Let $\Set{P}_m(\V{x}_{k})$ be the region occupied by measurement $m \in \{1, \dots, M\}$ that is defined as
\begin{equation}
    \Set{P}_m(\V{x}_{k}) = \{\V{p} \hspace{1mm} | \hspace{1mm} (m - 1) \ist T_{\mathrm{s}} \le \Vert \V{p} - \V{p}_k \Vert / c \le m \ist T_{\mathrm{s}} \} \nn
\end{equation}
with $T_{\mathrm{s}} = 1/ \Delta$. The birth \ac{pdf} of new \ac{pf} $n = N_{k - 1}^{(j)} \rmv+\rmv m$, is  $f_{\text{B}, n}^{(j)} (\V{\phi}_{k, n}^{(j)} | \V{x}_{k}) \rmv\propto\rmv f_{\text{B}}^{(j)} (\V{\phi} | \V{x}_{k})$ if $\V{p}_{k, n}^{(j)} \in \Set{P}_m(\V{x}_{k})$, and zero otherwise.
% \begin{equation}
%     f_{\text{B}, n}^{(j)}(\V{\phi} | \V{x}_{k}) \propto \begin{cases}
%         f_{\text{B}}^{(j)} (\V{\phi} | \V{x}_{k}), & \V{p} \in \Set{P}_m(\V{x}_{k}) \\
%         0, & \V{p} \not\in \Set{P}_m(\V{x}_{k})
%     \end{cases} \nn
% \end{equation}

To define the birth probability, we first note that the number of new features in cell $\Set{P}_m(\V{x}_k)$ is also Poisson distributed with mean $\mu_{\text{B}, n}^{(j)} = \mu_{\text{B}}^{(j)} \int \hspace{-1mm} \int_{\Set{P}_m(\V{x}_k)} f_{\text{B}}^{(j)} (\V{\phi} | \V{x}_{k}) \ist \mathrm{d} \V{p} \hspace{1mm} \mathrm{d} \gamma$. By assuming that there is at most one new feature in cell $\Set{P}_m(\V{x}_{k})$, we obtain $p_{\text{B}, n}^{(j)} = \mu_{\text{B}, n}^{(j)} / (\mu_{\text{B}, n}^{(j)} +1)$. The prior \ac{pdf} $f(\V{y}_{k, n}^{(j)} | \V{x}_{k}) = f(\V{\phi}_{k, n}^{(j)}, r_{k, n}^{(j)} | \V{x}_{k})$ for individual new \acp{pf} thus reads $f(\V{\phi}_{k, n}^{(j)}, 0 | \V{x}_{k}) = (1 - p_{\text{B}, n}^{(j)}) \ist f_{\text{D}}(\V{\phi}_{k, n}^{(j)})$ and $f(\V{\phi}_{k, n}^{(j)}, 1 | \V{x}_{k}) = p_{\text{B}, n}^{(j)} \ist f_{\text{B}, n}^{(j)} (\V{\phi}_{k, n}^{(j)} | \V{x}_{k})$. As a result of the Poisson point process assumption for newly appearing features, the new \ac{pf} can be assumed statistically\vspace{0mm} independent.

\subsection{The Factor Graph}

\begin{figure}[!tbp]
    \centering
    \psfrag{da1}[c][c][0.65]{\raisebox{-2mm}{\hspace{.8mm}$\V{y}_{1}$}}
    \psfrag{daI}[c][c][0.65]{\raisebox{-2.5mm}{\hspace{.3mm}$\V{y}_{\underline{N}}$}}
    \psfrag{db1}[c][c][0.65]{\raisebox{-2mm}{\hspace{.2mm}$\V{y}_{\scriptscriptstyle \underline{N} + 1}$}}
    \psfrag{dbJ}[c][c][0.65]{\raisebox{1mm}{$\V{y}_{N}$}}
    \psfrag{q1}[c][c][0.65]{\raisebox{-1mm}{$f_{1}$}}
    \psfrag{qI}[c][c][0.65]{\raisebox{-2mm}{$f_{\underline{N}}$}}
    \psfrag{v1}[c][c][0.65]{\raisebox{-2.3mm}{\hspace{.15mm}$f_{\scriptscriptstyle \underline{N} + 1}$}}
    \psfrag{vJ}[c][c][0.65]{\raisebox{-1mm}{\hspace{.3mm}$f_{N}$}}
    \psfrag{g1}[c][c][0.65]{\raisebox{-1mm}{\hspace{0mm}$f_{\V{z}}$}}
    \psfrag{ma11}[c][c][0.65]{\color{blue}{\raisebox{-0mm}{$\alpha_{1}$}}}
    \psfrag{ma12}[l][l][0.65]{\color{blue}{\raisebox{-2mm}{$\alpha_{1}$}}}
    \psfrag{maJ1}[c][c][0.65]{\color{blue}{$\alpha_{N}$}}
    \psfrag{maJ2}[l][l][0.65]{\color{blue}{$\alpha_{N}$}}
    \psfrag{mbJJ}[r][r][0.65]{\raisebox{-4mm}{\color{blue}{$\kappa_{N}$}}}
    \psfrag{mk11}[r][r][0.65]{\color{blue}{\raisebox{0.mm}{$\kappa_{1}$}}}
    \psfrag{mx11}[c][c][0.65]{\color{blue}{$\beta$}}
    \psfrag{mx12}[c][c][0.65]{\color{blue}{$\beta$}}
    \psfrag{mx13}[r][r][0.65]{\color{blue}{$\beta$}}
    \psfrag{mx2}[c][c][0.65]{\color{blue}{$\iota$}}
    \psfrag{dx}[c][c][0.65]{$\V{x}$}
    \psfrag{qx}[c][c][0.65]{$f$}
    \psfrag{de}[c][c][0.65]{$\V{C}_{\V{\epsilon}}$}
    \psfrag{qe}[c][c][0.65]{$f_{\V{\epsilon}}$}
    \psfrag{me1}[c][c][0.65]{\color{blue}{$\xi$}}
    \psfrag{me2}[c][c][0.65]{\color{blue}{$\xi$}}
    \psfrag{mn2}[c][c][0.65]{\color{blue}{$\nu$}}
    \includegraphics[scale=0.65]{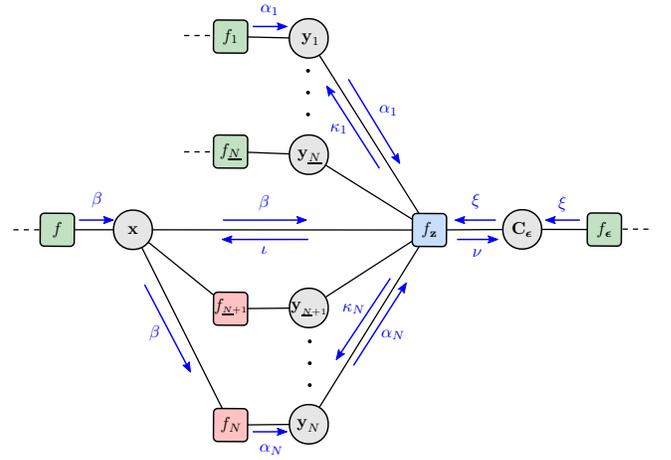}
    \vspace{2mm}
    \caption{Factor graph used for the development of the proposed direct \ac{slam} method. A single time step $k$ and a single \ac{pa} $j$ are shown. The indexes $k$ and $j$ are omitted and the following shorthand notations are used: $\underline{N} = N_{k - 1}^{(j)}$, $N = N_k^{(j)}$, $\V{x} = \V{x}_k$, $\V{y}_{n} = \V{y}_{k, n}^{(j)}$, $\M{C}_{\V{\epsilon}} = \M{C}_{\V{\epsilon}, k}^{(j)}$, $f_{\V{z}} = f(\V{z}_k^{(j)} | \V{x}_k, \V{y}_{k}^{(j)}, \M{C}_{\V{\epsilon}, k}^{(j)})$, $f = f(\V{x}_k | \V{x}_{k - 1})$, $f_{\V{\epsilon}} = f(\M{C}_{\V{\epsilon}, k}^{(j)} | \M{C}_{\V{\epsilon}, k - 1}^{(j)})$,  $f_n = f(\V{y}_{k, n}^{(j)} | \V{y}_{k - 1, n}^{(j)})$ for $n \in \{1, \dots, N_{k - 1}^{(j)}\}$, and $f_n = f(\V{y}_{k, n}^{(j)} | \V{x}_{k})$ for $n \in \{N_{k - 1}^{(j)} + 1, \dots, N_{k}^{(j)}\}$. Moreover, $\beta = \beta(\V{x}_k)$, $\alpha_n = \alpha(\V{y}_{k, n}^{(j)})$, $\xi = \xi(\M{C}_{\V{\epsilon}, k}^{(j)})$, $\iota = \iota(\V{x}_k ; \V{z}_k^{(j)})$, $\kappa_n = \kappa(\V{y}_{k, n}^{(j)}; \V{z}_k^{(j)})$, and $\nu = \nu(\M{C}_{\V{\epsilon}, k}^{(j)}; \V{z}_k^{(j)})$. }
    \label{fig:factor_graph}
   \vspace{-4mm}
\end{figure}

Based on the introduced statistical models and further common assumptions \cite{MeyKroWilLauHlaBraWin:J18}, the joint posterior \ac{pdf} $f(\V{x}_{0 : k}, \V{y}_{0 : k} | \V{z}_{1 : k})$ can then be factorized\vspace{-.5mm} as 
\begin{align}
    f(\V{x}_{0 : k}, \V{y}_{0 : k}, \M{C}_{\V{\epsilon}, 0 : k} | \V{z}_{1 : k}) \hspace{-35mm}& \nn \\
    &\propto f(\V{x}_0) \bigg( \prod^J_{j = 1} f(\M{C}_{\V{\epsilon}, 0}^{(j)}) \prod_{n = 1}^{N^{(j)}_0} f(\V{y}^{(j)}_{0, n} ) \bigg) \prod_{k^\prime = 1}^{k} f(\V{x}_{k'} | \V{x}_{k' - 1}) \nn \\
    & \times \prod^J_{j = 1} \bigg( \prod_{n = 1}^{N^{(j)}_{k^\prime - 1}} f(\V{y}^{(j)}_{k^\prime, n} | \V{y}^{(j)}_{k^\prime - 1, n}) \bigg) \bigg(  \prod_{n = N^{(j)}_{k^\prime - 1} + 1}^{N^{(j)}_{k^\prime}} \rmv\rmv\rmv f(\V{y}^{(j)}_{k^\prime\rmv, n} | \V{x}_{k'}) \rmv \bigg)  \nn \\
    & \times f(\M{C}_{\V{\epsilon}, k'}^{(j)} | \M{C}_{\V{\epsilon}, k' - 1}^{(j)})  f(\V{z}^{(j)}_{k'} | \V{x}_{k'}, \V{y}^{(j)}_{k'}, \M{C}_{\V{\epsilon}, k'}^{(j)})   
    \label{eq:factorization}
\end{align}
where we introduced $\V{x}_{0 : k} \triangleq [\V{x}_0^\T \cdots \V{x}_{k}^\T]^\T\rmv\rmv$, $\V{y}_{0 : k} \triangleq [\V{y}_0^\T$ $\cdots \V{y}_{k}^\T]^\T\rmv\rmv$, $\V{y}_{k} \triangleq [\V{y}_k^{(1) \T} \cdots \V{y}_k^{(J) \T} ]^\T\rmv\rmv$, $\M{C}_{\V{\epsilon}, k} \triangleq [\M{C}_{\V{\epsilon}, k}^{(1)} \cdots \M{C}_{\V{\epsilon}, k}^{(J)}]$, and $\M{C}_{\V{\epsilon}, 0 : k} \triangleq [\M{C}_{\V{\epsilon}, 0} \cdots \M{C}_{\V{\epsilon}, k}]$. Note that measurement vector $\V{z}_{1 : k}$ is here assumed observed and thus fixed. Based on the factorization in \eqref{eq:factorization}, $f(\V{x}_{0 : k}, \V{y}_{0 : k}, \M{C}_{\V{\epsilon}, 0 : k} | \V{z}_{1 : k})$ can be represented by the factor graph \cite{KscFreLoe:01} shown in Fig.~\ref{fig:factor_graph} for a single step and a single \ac{pa}.  Note that contrary to previous work \cite{LiaKroMey:J23}, the individual measurements $z_{k, m}^{(j)}$, $m \rmv\in\rmv \{1,\dots,M\}$ are coherent and thus not independent conditioned on $\V{x}_{k}$, $\V{y}^{(j)}_{k}$, and $\M{C}_{\V{\epsilon}, k}^{(j)}$. In contrast to \cite[Eq.~(2)]{LiaKroMey:J23}, it is thus not possible to factorize the joint likelihood function $f(\V{z}^{(j)}_{k} | \V{x}_{k}, \V{y}^{(j)}_{k}, \M{C}_{\V{\epsilon}, k}^{(j)})$ into individual likelihood functions for measurements $z_{k, m}^{(j)}$, $m \rmv\in\rmv \{1,\dots,M\}$.

% A \ac{bp} algorithm \cite{Loe:04} can be run on this factor graph to calculate the beliefs $\tilde{f}(\V{x}_{k})$ and $\tilde{f}(\V{y}_{k, n}^{(j)})$, which are approximations of the true marginal posterior $f(\V{x}_{k} | \V{z}_{1 : k})$ and $f(\V{y}_{k, n}^{(j)} | \V{z}_{1 : k})$, respectively. Note that the factor $f \big( \V{z}^{(j)}_{k} | \V{x}_k, \V{y}^{(j)}_{k}, \M{C}_{\V{\epsilon}}^{(j)} \big)$ depends on the agent and all \acp{pf} state and it is of high-dimension, for which the complexity of computing certain \ac{bp} messages grows exponentially with the number of \acp{pf}. To obtain a scalable BP method for direct multipath-based SLAM, we approximate these messages by Gaussian \acp{pdf}, the mean and covariance of which can be computed efficiently. From the beliefs $\tilde{f}(\V{x}_{k})$ and $\tilde{f}(\V{y}_{k, n}^{(j)})$, position estimates of agent and \acp{pf}, $\hat{\V{p}}_{k}$ and $\hat{\V{p}}_{k, n}^{(j)}$, can finally be computed.
% \vspace{-1mm} 

%-------------------------------------------------------------------------------------

%-------------------------------------------------------------------------------------

%-------------------------------------------------------------------------------------

%-------------------------------------------------------------------------------------
\section{The Proposed \acf{bp} Method} \label{sec:bp}
\vspace{0mm}

This section introduces the proposed \ac{bp} method for direct \ac{slam}. \ac{bp} is an efficient technique to compute approximate marginal posterior \acp{pdf}, also known as beliefs, in terms of the sum-product rule by performing local operations, so-called ``messages'', passed along the edges of a factor graph. Since the factor graph of the considered direct \ac{slam} problem has loops, there is no fixed message passing order  \cite{Loe:04,KscFreLoe:01}. We apply the following message passing schedule: (i) messages are only passed forward in time, (ii) along the edges connecting an agent state variable node ``$\V{x}_{k}$'' and a new \ac{pf} state variable node ``$\V{y}_{k, n}^{(j)}$'', messages are only sent from the former to the latter\vspace{0mm} \cite{LeiMeyHlaWitTufWin:J19}.

\subsection{\Ac{bp}  Message Passing} \label{subsec:bp_bp}

Following the sum-product message passing rules, we first compute the prediction messages $\beta(\V{x}_k)$, $\xi(\M{C}_{\V{\epsilon}, k}^{(j)})$, and $\alpha(\V{y}_{k, n}^{(j)})$, for $n \in \{1, \dots, N_{k - 1}^{(j)} \}$. These messages are obtained based on beliefs from the previous time step and the state-transition models introduced in Section \ref{sec:models},\vspace{-.3mm} i.e.,
\begin{align}
    \beta(\V{x}_k) &= \int f(\V{x}_k | \V{x}_{k - 1}) \tilde{f}(\V{x}_{k - 1}) \hspace{1mm} \mathrm{d} \V{x}_{k - 1} \nn \\[1.8mm]
    \xi(\M{C}_{\V{\epsilon}, k}^{(j)}) &= \int f(\M{C}_{\V{\epsilon}, k}^{(j)} | \M{C}_{\V{\epsilon}, k - 1}^{(j)}) \tilde{f}(\M{C}_{\V{\epsilon}, k - 1}^{(j)}) \hspace{1mm} \mathrm{d} \M{C}_{\V{\epsilon}, k - 1}^{(j)} \nn \\[1.8mm]
    \alpha(\V{y}_{k, n}^{(j)}) &= \sum_{\V{y}_{k - 1, n}^{(j)}} f(\V{y}_{k, n}^{(j)} |\V{y}_{k - 1, n}^{(j)}) \tilde{f}(\V{y}_{k - 1, n}^{(j)}) \nn \\[-7mm]
   \nonumber
\end{align}
where $\sum_{\V{y}_{k - 1, n}^{(j)}}\vspace{-.8mm}$ denotes ``marginalizing out'' $\V{y}_{k - 1, n}^{(j)}$. This marginalization includes the summation over $r_{k - 1, n}^{(j)}$ and the integration over $\V{p}_{k - 1, n}^{(j)}$ and $\gamma_{k - 1, n}^{(j)}$. Here, $\tilde{f}(\V{x}_{k - 1})$, $\tilde{f}(\M{C}_{\V{\epsilon}, k - 1}^{(j)})$, and $\tilde{f}(\V{y}_{k - 1, n}^{(j)})$ are the beliefs computed at the previous time step. For the new \acp{pf} $n \in \{N_{k - 1}^{(j)} + 1, \dots, N_{k}^{(j)} \}$, since we only pass from ``$\V{x}_{k}$'' to ``$\V{y}_{k, n}^{(j)}$'', the corresponding ``birth messages'' are given by $\alpha(\V{y}_{k, n}^{(j)}) = \int f(\V{y}_{k, n}^{(j)} | \V{x}_{k}) \beta(\V{x}_{k}) \hspace{1mm} \mathrm{d} \V{x}_{k}.$
% \begin{equation}
%     \alpha(\V{y}_{k, n}^{(j)}) = \int f(\V{y}_{k, n}^{(j)} | \V{x}_{k}) \beta(\V{x}_{k}) \hspace{1mm} \mathrm{d} \V{x}_{k}. \nn
% \end{equation}

Furthermore, based on sum-product message passing rules, the measurement update messages $\iota(\V{x}_k ; \V{z}_k^{(j)})$, $\nu(\M{C}_{\V{\epsilon}, k}^{(j)}; \V{z}_k^{(j)})$, and $\kappa(\V{y}_{k, n}^{(j)}; \V{z}_k^{(j)})$ that introduce the information of the current measurement $\V{z}_{k}^{(j)}$, can be obtained\vspace{1mm} as
\begin{align}
    \iota(\V{x}_k; \V{z}_k^{(j)} ) = & \sum_{\V{y}_{k}^{(j)}} \int  f(\V{z}_{k}^{(j)} | \V{x}_k, \V{y}_{k}^{(j)}, \M{C}_{\V{\epsilon}, k}^{(j)}) \xi(\M{C}_{\V{\epsilon}, k}^{(j)})  \nn \\[-1.3mm]
    & \times \prod_{n = 1}^{N_k^{(j)}} \alpha(\V{y}^{(j)}_{k, n}) \hspace{1mm} \mathrm{d} \M{C}_{\V{\epsilon}, k}^{(j)} \label{eq:bp_iota} \\[4mm]
    \nu(\M{C}_{\V{\epsilon}, k}^{(j)}; \V{z}_k^{(j)} ) = & \sum_{\V{y}_{k}^{(j)} } \int f(\V{z}_{k}^{(j)} | \V{x}_k, \V{y}_{k}^{(j)}, \M{C}_{\V{\epsilon}, k}^{(j)}) \beta(\V{x}_k) \nn \\[-1.3mm]
    & \times \prod_{n = 1}^{N_k^{(j)}} \alpha(\V{y}^{(j)}_{k, n}) \hspace{1mm} \mathrm{d} \V{x}_k. \label{eq:bp_nu} \\[4mm]
    \kappa(\V{y}_{k, n}^{(j)}; \V{z}_{k}^{(j)} ) = & \sum_{\V{y}_k^{(j)} \backslash \V{y}_{k, n}^{(j)}} \int \hspace{-2mm} \int f(\V{z}_k^{(j)} | \V{x}_k, \V{y}_k^{(j)}, \M{C}_{\V{\epsilon}, k}^{(j)}) \xi(\M{C}_{\V{\epsilon}, k}^{(j)}) \nn \\[-1.3mm]
    & \times \beta(\V{x}_k)  \prod_{\substack{n' = 1 \\ n' \ne n}}^{N_k^{(j)}} \alpha(\V{y}^{(j)}_{k, n'}) \hspace{1mm} \mathrm{d} \M{C}_{\V{\epsilon}, k}^{(j)} \hspace{1mm} \mathrm{d} \V{x}_k \label{eq:bp_kappa} \\[-4.5mm]
    \nonumber
\end{align}
where $\sum_{\V{y}_k^{(j)} \backslash \V{y}_{k, n}^{(j)}}$ denotes\vspace{-.8mm} ``marginalizing out'' all elements of $\V{y}_k^{(j)}$ except for $\V{y}_{k, n}^{(j)}$. Note that the messages in \eqref{eq:bp_iota}-\eqref{eq:bp_kappa} all involve ``marginalizing out''  $\V{y}_k^{(j)}$, the dimension of which grows linearly with the number of \acp{pf}. The complexity of this operation scales exponentially with the number of \acp{pf}. We observe that, due to the functional form of $ f(\V{z}_k^{(j)} | \V{x}_k, \V{y}_k^{(j)}, \M{C}_{\V{\epsilon}, k}^{(j)})$, these measurement update messages are a mixture of zero-mean complex Gaussian \acp{pdf} of $\V{z}_k^{(j)}\rmv\rmv$. To avoid the high computational complexity, we approximate each of these messages by a single zero-mean complex Gaussian \ac{pdf} of $\V{z}_k^{(j)}$ \cite{LiaKroMey:J23}. In particular, we calculated the following approximate messages: $\tilde{\iota} (\V{x}_k; \V{z}_k^{(j)} ) = \mathcal{CN}(\V{z}_k^{(j)}; \V{0}, \M{C}_{\iota, k}^{(j)}),$
$\tilde{\nu} (\M{C}_{\V{\epsilon}, k}^{(j)}; \V{z}_k^{(j)} ) = \mathcal{CN}(\V{z}_k^{(j)}; \V{0}, \M{C}_{\nu, k}^{(j)})$, and $\tilde{\kappa} (\V{y}_{k, n}^{(j)}; \V{z}_{k}^{(j)} ) = \mathcal{CN}(\V{z}_k^{(j)}; \V{0}, \M{C}_{\kappa, k, n}^{(j)})$. The covariance matrices $\M{C}_{\iota, k}^{(j)}$, $\M{C}_{\nu, k}^{(j)}$, and $\M{C}_{\kappa, k, n}^{(j)}$ are computed via moment matching, i.e., they are the covariance of the original \ac{bp} messages in \eqref{eq:bp_iota}-\eqref{eq:bp_kappa}. The computational complexity of calculating these covariance matrices scales only linearly with the number \acp{pf} (see \cite{LiaKroMey:J23} for details).
% \begin{align}

\subsection{Belief Calculation, State Declaration and Estimation} \label{subsec:bp_state}
With the \ac{bp} messages computed as discussed in Sec.~\ref{subsec:bp_bp}, beliefs can be obtained\vspace{-1mm} as
\begin{align}
    \tilde{f}(\V{x}_k) &\propto \beta(\V{x}_k) \prod^{J}_{j = 1} \tilde{\iota}(\V{x}_k; \V{z}_k^{(j)} ) \nn \\[2mm]
    \tilde{f}(\V{y}_{k, n}^{(j)}) &\propto \alpha(\V{y}_{k, n}^{(j)}) \tilde{\kappa}(\V{y}_{k, n}^{(j)}; \V{z}_k^{(j)} ) \nn \\[4mm]
    \tilde{f}(\M{C}_{\V{\epsilon}, k}^{(j)}) &\propto  \xi(\M{C}_{\V{\epsilon}, k}^{(j)}) \tilde{\nu}(\M{C}_{\V{\epsilon}, k}^{(j)}; \V{z}_k^{(j)} ). \\[-3.5mm]
    \nn
\end{align}
These beliefs are computed using particles by following the importance sampling principle, with $\beta(\V{x}_k)$, $ \alpha(\V{y}_{k, n}^{(j)})$, and $ \xi(\M{C}_{\V{\epsilon}, k}^{(j)}) $, respectively, used as the proposal PDFs \cite{AruMasGorCla:02}. To determine the number of \acp{pf} at each time step, a \ac{pf} is declared to exist \vspace{-.5mm} if its existence probability $\tilde{f}(r_{k, n}^{(j)} = 1)$, which computed from $\tilde{f}(\V{y}_{k, n}^{(j)})$, is larger than a threshold $T_{\text{dec}}$. 

The agent and existing \ac{pf} positions are finally estimated based on \ac{mmse} estimation \cite{Kay:B93}, i.e.\vspace{-1.5mm},
\begin{align}
    \hat{\V{p}}_{k}^{} &= \int \V{p}_{k} \tilde{f}(\V{p}_{k}) \hspace{1mm} \mathrm{d} \V{x}_{k} \nn \\[2mm]
    \hat{\V{p}}_{k, n}^{(j)} &= \int \V{p}_{k, n}^{(j)} \tilde{f}(\V{p}_{k, n}^{(j)} | r_{k, n}^{(j)} = 1) \hspace{1mm} \mathrm{d} \V{p}_{k, n}^{(j)} \nn \\[-4mm]
    \nn
\end{align}
where $\tilde{f}(\V{p}_{k})$ and $\tilde{f}(\V{p}_{k, n}^{(j)} | r_{k, n}^{(j)} = 1)$ can\vspace{-.5mm} be obtained by marginalizing from $\tilde{f}(\V{x}_k)$ and $\tilde{f}(\V{y}_{k, n}^{(j)})$, respectively. 

The number of \acp{pf} increases with time. To limit computational complexity, we prune \acp{pf}, i.e., remove them from the state space if their existence probability is smaller than\vspace{2mm} $T_{\text{pru}}$.

\begin{figure}[!tbp]
    \centering
    \hspace{-4mm}
    \resizebox{0.9\linewidth}{!}{\input{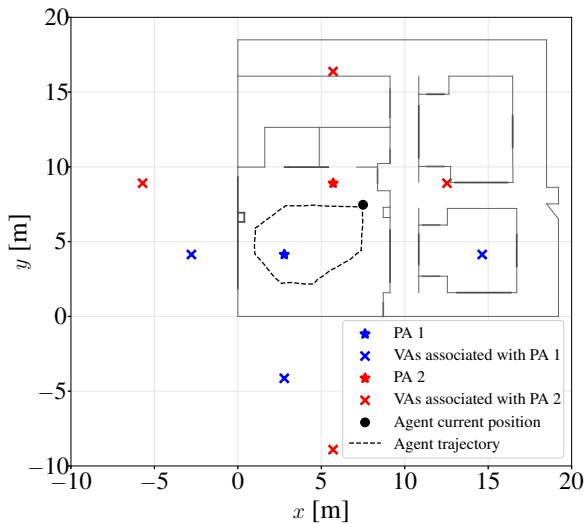}}
    \caption{Floor plan used for simulation. Agent and \acp{va} positions for time step $k = 1$ are shown.}
    \label{fig:floorplan}
    \vspace{-4mm}
\end{figure}

%-------------------------------------------------------------------------------------

%-------------------------------------------------------------------------------------
\section{Simulation Results}

We consider a $2$-D indoor localization scenario with $J = 2$ \acp{pa}. The floor plan, positions of \acp{pa}, and the agent's trajectory with $679$ time steps are depicted in Fig.~\ref{fig:floorplan}. The number propagation paths and their corresponding \acp{va} are computed based on ray tracing \cite{Bor:J84, Ant:B92}. At each time step, the radio signal received by each \ac{pa} is generated following \eqref{eq:meas_model}. Only propagation paths and corresponding VAs that reflect from a single surface are considered. We set $M = 41$ and $\Delta = 10\,$MHz, which corresponds to $400\,$MHz signal bandwidth and a maximum range of 30m. In total, $100$ simulations are performed. We assume that the measurement noise $\V{\epsilon}_k^{(j)}$ is \ac{iid} across $m \in \{1, \dots, M\}$, leading to $\M{C}_{\V{\epsilon}, k}^{(j)} = \sigma_{\V{\epsilon}, k}^{(j) 2} \M{I}_M$. The state transition \acp{pdf} $f(\V{x}_{k} | \V{x}_{k - 1})$, $f(\V{\phi}_{k, n}^{(j)} | \V{\phi}_{k, n - 1}^{(j)})$, and $f(\sigma_{\V{\epsilon}, k}^{(j) 2} | \sigma_{\V{\epsilon}, k - 1}^{(j) 2})$ follow a constant-velocity model $\V{x}_k = \M{F}\V{x}_{k - 1} + \M{W}\V{q}_{\V{x}, k}$ \cite[Ch. 4]{ShaKirLi:B02}, random walk model $\V{\phi}_k^{(j)} = \V{\phi}_{k - 1}^{(j)} + \V{q}_{\V{\phi}, k}^{(j)}$, and a Gamma distribution $\mathcal{G}(\sigma_{\V{\epsilon}, k}^{(j) 2}; \sigma_{\V{\epsilon}, k - 1}^{(j) 2} / c_{\V{\epsilon}}, c_{\V{\epsilon}})$, respectively. The covariance of $\V{q}_{\V{x}, k}$ and $\V{q}_{\V{\phi}, k}^{(j)}$ are set to $\M{\Sigma}_{\V{q}, \V{x}} = 10^{-4}\M{I}_2$ and $\M{\Sigma}_{\V{q}, \V{\phi}} = \text{diag}(10^{-8}, 10^{-8}, 10^{-4})$, and we set $c_{\V{\epsilon}} = 10$. We set the declaration threshold to $T_{\text{dec}} = 0.5$, the pruning threshold to $T_{\text{pru}} = 10^{-2}$, the survival probability to $p_{\text{s}} = 0.999$, and the birth probability to $p_{\text{B}, n}^{(j)} = 10^{-4}$.

\begin{figure}[!tbp]
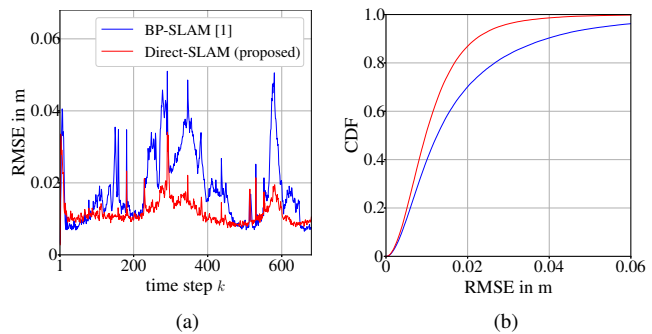

    \centering
    % \subfloat[]{\resizebox{0.48\linewidth}{!}{\input{Figs/gospa.pgf}}\label{fig:performance_gospa}} \hspace{1mm}
    \hspace{-4mm}
    \subfloat[\hspace{-10mm} ]{\resizebox{0.48\linewidth}{!}{\input{Figs/rmse.pgf} }\label{fig:performance_rmse}} \hspace{1mm}
    \subfloat[\hspace{-4mm} ]{\resizebox{0.48\linewidth}{!}{\input{Figs/cdf.pgf}}\label{fig:performance_cdf} }
    \caption{(a) RMSE of the estimated agent position versus time step $k$ over 100 simulation runs and (b) empirical CDFs of the RMSEs.}
    \label{fig:performance_localization}
    \vspace{-1mm}
\end{figure}

\begin{figure}[!t]
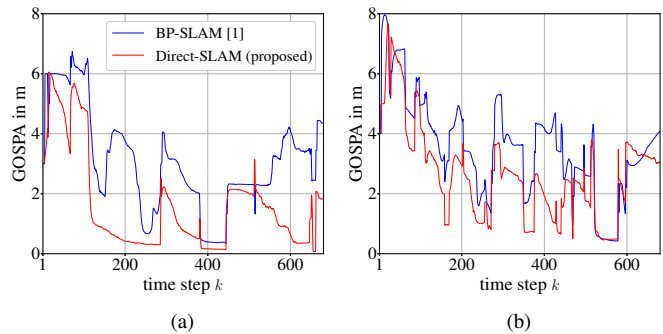

    \centering
    \subfloat[\hspace{-6mm} ]{\resizebox{0.48\linewidth}{!}{\input{Figs/gospa_sensor1.pgf}}} \hspace{1mm}
    \subfloat[\hspace{-6mm} ]{\resizebox{0.48\linewidth}{!}{\input{Figs/gospa_sensor2.pgf}}}
    \caption{GOSPA error of estimated PFs associated with (a) PA 1 and (b) PA 2 averaged over 100 simulation runs.}
    \label{fig:performance_mapping}
\end{figure}

We compare our proposed ``Direct-SLAM'' method with the state-of-the-art ``BP-SLAM'' method \cite{LeiMeyHlaWitTufWin:J19}. The reference method uses the \ac{mpc} estimates
from a snapshot-based parametric sparse Bayesian learning channel estimator \cite{ShuWanJos:13,GerMecChrXenNan:J16,HanFleuRao:J18,GreLeiWitFle:J23} as measurements. The reference method uses the particle-based nonparametric \ac{bp} method in \cite{MeyHliHla:J16,MeyBraWilHla:J17}. Fig.~\ref{fig:performance_rmse} shows the \ac{rmse} of the agent position, and Fig.~\ref{fig:performance_cdf} shows their empirical \acp{cdf}. It can be seen that Direct-SLAM significantly outperforms BP-SLAM. The large \ac{rmse} of the agent position, e.g., around $k \in [200, 400]$, is related to a challenging geometry where multiple \acp{pf} have a similar propagation delay. As a result of this geometry, the channel estimator needed for BP-SLAM cannot accurately extract \acp{mpc}, i.e., due to finite resolution capabilities limited by signal bandwidth, fewer \acp{mpc} than actual signal components are extracted. The performance of the proposed Direct-SLAM method, which does not rely on a channel estimator, suffers less severely in this challenging geometry. Fig.~\ref{fig:performance_mapping} shows the \ac{gospa} error \cite{RahGarSve:17} of \acp{pf} with cutoff parameter $c = 2$, order $p = 1$, indicating that Direct-SLAM also achieves superior mapping accuracy compared to\vspace{0mm} BP-SLAM. 

\acresetall
\section{Conclusion} \label{sec:conclusion}
We propose a \ac{bp}-based method for multipath-based \ac{slam} that uses received radio signals as measurements. By avoiding the use of a channel estimator as a preprocessing stage, the proposed approach can better exploit location information in received radio signals and thus succeed in geometrically challenging environments. For direct multipath-based \ac{slam}, we introduced a new statistical model to describe the data-generating process of received radio signals, combining the Swerling 1 model for correlated measurements and a Bernoulli existence model. A factor graph is constructed based on the new statistical model. This factor graph provides the blueprint for developing an efficient \ac{bp} method for direct multipath-based \ac{slam}. For an accurate approximation, some of the \ac{bp} messages are represented by random samples ``particles'' and others by a mean and covariance matrix obtained via moment matching. Performance evaluation is conducted based on synthetic data in a realistic scenario. It is demonstrated that the proposed method outperforms a state-of-the-art conventional method that relies on preprocessing of the received radio signal using a snapshot-based channel estimator. Future research avenues include \ac{bp}-based processing that is neural enhanced \cite{LiaMey:J23} or has an embedded particle flow \cite{LiyDau:J23,ZhaMey:J23,JanMeySnyWigBauHil:J23,WieLeiMeyWit:TSIPN2023} as well as a representation of environmental features that enables data fusion \cite{LeiVenTeaMey:J23}.

\section*{Acknowledgement} \label{sec:acknowledgement}

The material presented in this work was supported by the Under Secretary of Defense for Research and Engineering under Air Force Contract No. FA8702-15-D-0001 and by Qualcomm Innovation Fellowship No. 492866.

\pagebreak

% \renewcommand{\baselinestretch}{1}
% \selectfont
\bibliographystyle{ieeetr}
\bibliography{IEEEabrv,StringDefinitions,Papers,Books,ref,refBooks}

%---------------------------------------------------------------------------%
%                                   cover sheet                             %
%---------------------------------------------------------------------------%
% \newpage
% {
% \onecolumn % force the text after this point to be 1 column

% \title{\paperTitle}
% \author{Authors\\
%   \vspace{0.5cm}
%   \textcolor{BLUE}\versionDT
% }
% \maketitle

% \tableofcontents

% %% ------ Catechism  ------ %%

% \newpage

% \input{LatexInclusion/catechism} 

% \bigskip
% Helpful hints:

% \vfill
% \noindent \textit{Template version \today}
% }

\end{document}